# LOW FIELD MAGNETOTRANSPORT IN STRAINED Si/SiGe CAVITIES

**G. Scappucci, L. Di Gaspare, and F. Evangelisti**
Dipartimento di Fisica "E. Amaldi", Università di Roma TRE, V. Vasca Navale 84, 00146 Roma, Italy

**E. Giovine, A. Notargiacomo, and R. Leoni**
Istituto di Fotonica e Nanotecnologie, IFN-CNR, V. Cineto Romano 42, 00156 Roma, Italy

**V. Piazza, P. Pingue, and F. Beltram**
NEST-INFM and Scuola Normale Superiore, Via della Faggiola 19, 56126 Pisa, Italy

## Abstract

Low field magnetotransport revealing signatures of ballistic transport effects in strained Si/SiGe cavities is investigated. We fabricated strained Si/SiGe cavities by confining a high mobility Si/SiGe 2DEG in a bended nanowire geometry defined by electron-beam lithography and reactive ion etching. The main features observed in the low temperature magnetoresistance curves are the presence of a zero-field magnetoresistance peak and of an oscillatory structure at low fields. By adopting a simple geometrical model we explain the oscillatory structure in terms of electron magnetic focusing. A detailed examination of the zero-field peak lineshape clearly shows deviations from the predictions of ballistic weak localization theory.

## I. INTRODUCTION

The magnetoresistance of nanostructures at low magnetic fields is related to electron trajectories and can provide, therefore, useful information on device geometry and field distribution. Relevant effects are magnetic focusing (MF) and weak localization (WL). The two effects become closely linked in ballistic nanostructures since in ballistic motion also the zero magnetic field peak, which is a signature of weak localization in diffusive transport, is related to the electron paths in the devices.

First experimental observations and theoretical studies on electron focusing in mesoscopic solid state devices were reported by van Houten and collaborators[1] shortly after the discovery of conductance quantization in two dimensional electron gas (2DEG) quantum point contacts.[2] In that work a transverse electron focusing geometry was used, in which a perpendicular magnetic field bends the trajectory of ballistic electrons between two quantum point contacts, defined by split-gates on a high mobility GaAs/AlGaAs 2DEG. From then on, transverse electron focusing experiments have led to significant achievements such as the detection of composite fermions in the fractional quantum Hall regime[3,4] or, more recently, the detection of spin polarized currents.[5,6] Internal MF was mostly studied in cavities (or arrays of cavities) of different shape obtained by split-gate techniques[7-10] or – in only one case reported in literature – by dry etching techniques[11] on GaAs/AlGaAs 2DEG high mobility heterostructures. To the best of our knowledge, no studies of this kind have been reported in Si/SiGe structures as yet.

In the magnetoresistance curves of ballistic quantum dots or cavities a peak is usually present at zero magnetic field. This feature, which is similar to the WL peak found in diffusive transport has been the subject of a great interest recently. Ballistic quantum dots or cavities are systems essentially free of bulk disorder so that the constructive electron interference process giving rise to a WL phenomenon is associated with the backscattered electron trajectories determined by scattering at the cavities boundaries. The peak lineshape is expected to be sensitive to the shape of the cavity, while in diffusive WL it reflects the disorder of the system. Moreover, it was predicted[12,13] that the WL lineshape could be used to obtain information on the dynamics of the carriers in the cavity, a chaotic carrier dynamics – i.e. unstable for infinitesimal variations of the initial conditions – leading to a Lorentzian-like negative magnetoresistance peak, while a regular dynamics – i.e. stable for infinitesimal variations of the initial conditions – leading to a linear lineshape.

From the experimental point of view the situation is not clear. Some WL studies on differently shaped cavities, supporting different carrier dynamics, have confirmed the expected difference in the negative magnetoresistance lineshape.[14] However, this behavior was not found in other works. In circular cavities, expected to exhibit a regular dynamics, a Lorentzian lineshape instead of the linear one was found.[15,16] Moreover, a transition between different lineshapes was observed in the same structure upon varying the gate voltage.[17] These and others experimental observations have raised the question whether the WL lineshape analysis is a reliable indicator of the carrier dynamics in the cavity and whether the zero-field magnetoresistance peak should be related to a WL-like phenomenon or could be also explained in other frameworks. Simulations by Akis and co-workers[18,19] have shown that the zero-field peak and the shoulders that develop on both sides of it can be explained in terms of conductance resonances reflecting the intrinsic energy spectrum of the ballistic quantum dot. In a successive paper the same group has compared in detail experimental data and simulations and interpreted the magnetoresistance structures as a magnetospectroscopy of the quantum dot intrinsic density of states.[20] A review of related theoretical and experimental issues can be found in a recent report.[21]

In the present paper we address experimentally this subject by investigating magnetotransport in cavities fabricated from Si/SiGe two dimensional electron gas. To our knowledge, Si/SiGe-based ballistic cavities are investigated for the first time. The adopted geometry for the cavities allows us to change the configuration from open to isolated quantum dot by varying the bias of a control gate placed on top of the cavity itself. Therefore, the same device can be used as a single electron transistor.[22]

## II. PROPERTIES OF THE Si/SiGe 2DEG

The high-mobility two-dimensional electron gas was grown by low-pressure chemical vapour deposition in a UHV chamber with a base pressure of $10^{-10}$ Torr, using silane and germane. The 2DEG was obtained by depositing on a SiGe virtual substrate the following layers: *i)* a tensile Si channel layer (thickness ~11 nm); *ii)* a $Si_{0.81}Ge_{0.19}$ spacer layer (thickness ~11 nm) and *iii)* an *n*-doped $Si_{0.81}Ge_{0.19}$ layer (thickness ~11.5 nm). The structures were completed by a second 35-nm-thick $Si_{0.81}Ge_{0.19}$ spacer layer followed by a final 15-nm-thick Si cap layer. Details of the properties of the 2DEG samples can be found elsewhere.[23,24]

The transport parameters of the 2DEGs were extracted from a standard analysis of the low-field magnetoresistance measurements performed on mesa-etched Hall bars. In Fig. 1(a) we report the low field longitudinal magnetoresistivity $\rho_{xx}(B)$ as measured at T=250mK with standard low frequency lock-in techniques in a four-terminal configuration. A small weak localization (WL) zero-field peak develops in a relatively narrow field range (~10mT), after which the resistivity saturates to a value of $\rho_{xx}(B \gg B_C)=122\Omega/\square$. $B_C$ is the characteristic field for the suppression of weak localization. For fields larger than 0.25 T well resolved Shubnikov-de Haas (SdH) oscillations appear. The 2DEG carrier density was estimated from the periodicity of the SdH oscillations versus the inverse of the magnetic field. In Fig. 1(b) we report the Fourier spectrum of $\Delta\rho_{xx}(B)=\rho_{xx}(B)-\rho_{xx}(B \gg B_C)$ – shown as a function of 1/B in the inset of the same figure – exhibiting a single peak from which a carrier concentration $n_{2D}=7.62\times10^{11}cm^{-2}$ is estimated. From the carrier concentration $n_{2D}$ and the classical conductivity of the 2DEG $\sigma=1/\rho_{xx}(B \gg B_C)$ the evaluation of the other parameters relative to diffusive transport follows directly: we found the electronic mobility to be $\mu_E$ =6.7x10$^4$ cm$^2$/Vs, the momentum relaxation time $\tau_E$ = 7.3 ps and the mean free path of the 2D unconstrained carriers $\lambda_E$ =0.7 μm. Other parameters such as diffusion coefficient $D$ and Fermi velocity $v_F$ are reported in Table I. In order to evaluate the single particle collision time $\tau_q$ and so gain insights into the scattering mechanisms limiting the electronic mobility we performed an analysis of the amplitude of the SdH oscillations according to the procedures described in Ref. 25. The Dingle plot of our data is reported in Fig. 1(c). The relative amplitude of the SdH oscillations $\Delta\rho_{xx}/\rho_0$, corrected by a damping factor $\chi(T,\omega_C) = (2\pi^2 kT/\hbar\omega_C)/\sinh(2\pi^2 kT/\hbar\omega_C)$ is reported in a logarithmic scale as a function of 1/B. $\omega_C = eB/m_e^*$ is the cyclotron frequency and $m_e^*$ the effective mass for electrons in the Si/SiGe 2DEG. In calculating the thermal damping factor we assumed a temperature T=250 mK and an effective mass of the 2DEG carriers $m_e^*$=0.19 $m_e$, with $m_e$ electronic mass. Since theory predicts for the amplitude of the SdH oscillations the relation $\Delta\rho_{xx} = 4\rho_0 \chi(T) \exp(-\pi/\omega_C \tau_q)$, a Dingle plot is expected to give a straight line whose slope is related to the single particle lifetime $\tau_q$ and an intercept at 1/B=0 that reaches the value of 4. A linear fit to the experimental data is reported in Fig. 1(c) from which we estimate $\tau_q$= 2.3 ps, so that the ratio $\tau_E/\tau_q$ between the total scattering rate and the momentum relaxation rate is found to be 3.2. This value indicates that the main source of scattering are charged impurities, but its relatively small absolute value suggests that there is still a short range contribution to the scattering.[26] As for the intercept at 1/B=0, we found a value of 2.19 which is smaller than the theoretical prediction. Deviations in the expected value of the Dingle plot intercept are reported frequently in the literature. Possible explanations include, among the others, the presence of a parallel conducting path in the heterostructure containing the 2DEG.[27]

Finally, we estimated the phase coherence length $L_\Phi$ and the phase coherence time $\tau_\Phi$ by evaluating the characteristic field $B_C$ taken as the field at which the weak localization correction to the conductivity is reduced by a factor of two. From the magnetoresistance curve we found $B_C \sim 3$ mT. In 2D the critical field $B_C$ is related to the phase coherence length by the relation $B_C \approx \hbar/e L_\Phi^2$ so that we estimate $L_\Phi \sim 0.5$ μm and, from $L_\Phi = \sqrt{D\tau_\Phi}$, the phase coherence time $\tau_\Phi \sim 7$ ps. Similar values were found in previous weak localization studies on $n$-type Si/SiGe heterostructures.[28,29]

## III. THE Si/SiGe CAVITIES

### A. Fabrication and experimental

The devices were fabricated by confining the 2DEG in a bended nanowire geometry defined by electron-beam lithography (EBL) and reactive ion etching. In Fig. 2 we report the scanning electron micrograph of a device. The cavities were obtained by the lateral displacement of a central 550-nm-long segment of 2-μm-long and 250-nm-wide nanowires, each defined in the inner region of a mesa structure confining the 2DEG. The mesa are shaped so as to allow electrical characterization of the devices both in a two and in a four-terminal configuration. Nanowires with an increasing lateral shift of the central section were fabricated on the same 2DEG sample. In this work we will mainly discuss the transport properties of devices characterized by a 180 nm shift that results in two constrictions with a geometrical width of 70 nm, connecting the central cavity with the other two segments of the wire. The devices were completed by depositing a 25-nm-thick $SiO_2$ layer by Electron-Cyclotron-Resonance

Plasma-Enhanced Chemical-Vapour-Deposition and defining by EBL and lift-off a central 500-nm-wide aluminium control gate on top of the cavity (see Fig. 2).

Due to sidewall depletion caused by the fabrication process, the constrictions have an effective width smaller than the lithographic one and behave as quantum point contacts connecting the cavity to the source and drain. Since the cavity dimensions (nominally ~550 nm long and 250 nm wide) are smaller than the mean free path, the transport in the cavities is expected to show phenomena typical of ballistic regime.

Electrical characterization of the devices was performed in a dilution refrigerator by measuring the conductance and the magnetoresistance in the 50 mK-4.2 K temperature range using standard AC low-frequency lock-in techniques. Transport measurements were performed both in two and four-terminal configuration, keeping voltage and current through the device small enough to prevent electron heating.

In all the measurements discussed in the present paper the gate bias was such that the transport through the cavity was studied in an "open" regime.

In Fig. 3 we report the low field magnetoresistance measured at T=50 mK in the four-terminal configuration ($R_{2,3}$, thicker line) and in the two-terminal configuration ($R_{1,4}$, thinner line) as well as a two-terminal "background" resistance ($R_{1,2}$, dotted line) that is indicative of transport through the 2DEG mesa region nearby the nanowire and the cavity. Refer to the inset in Fig. 3 for contact schematics.

In the curves $R_{2,3}(B)$ and $R_{1,4}(B)$, relative to transport through the cavity, the two most noteworthy features are the presence of a zero-field magnetoresistance peak and of an oscillatory structure at low fields (highlighted by arrows in Fig. 3) distinct from Shubnikov-de Haas oscillation that develop for B>~0.45 T.

We will first discuss the oscillations that dominate the magnetoresistance in the low magnetic-field range and next we will examine the zero-field peak.

**B. Magnetic focusing**

The absence of oscillations in the 0.1-0.45 T range of the $R_{1,2}$ curve in Fig. 3 points out that the oscillatory structure originates in the transport through the cavity. Moreover, the observation that the $R_{2,3}(B)$ and $R_{1,4}(B)$ maxima and minima were fully reproducible upon magnetic field or temperature cycling suggests the presence of a robust effect. We interpret this oscillatory structure in terms of internal magnetic focusing due to the commensurability between the ballistic trajectories of electrons in the cavity and the cavity geometry.

In a simple but useful picture, the ballistic motion of an electron in a cavity can be described as the motion of a charged particle in a billiard whose trajectory, given the initial conditions (Fermi velocity $v_F$ and entrance angle $\phi$), depends on the magnetic field B through the Lorentz force and the scattering with the billiard walls. Among the possible and complex paths, the cavity geometry (shape and opening positions) determines the existence of special trajectories that permit electrons to be backscattered to the billiard entrance or transmitted to the exit after a limited number of wall collisions. The magnetic fields that determine the trajectories of fully reflected or easily transmitted electrons are expected to be those at which the magnetoresistance curve develops, respectively, maxima and minima. Adopting similar internal focusing models, in most of the reported studies it was possible to correlate unambiguously the oscillatory magnetoresistance structure with specific electron trajectories.[30,31]

As for our case, we report in Fig. 4 the schematic of a rectangular cavity (dimensions L x D) and the specific trajectories that allow an electron entering the cavity with an entrance angle $\phi$ on the lower left corner to leave it after $N=0$, 1 and 2 collisions with the lower wall (i.e. $N+1$ collision with the upper wall).

The commensurability conditions giving rise to enhanced transmission are met when electrons first collide with the lower wall at a fraction $1/j$ of its length $L$, with $j = 1, 2, 3,...$ index related to $N$ by the relation $j = N+1$. The study of the cyclotron motion in the cavity gives the equations for the trajectories and allows us to find the relationship that determines the magnetic field $B_j$ for the trajectories of enhanced transmission, as a function of the geometrical parameters of the cavity (L, D, $\phi$) and the physical properties of the 2DEG ($v_F$, $m_e^*$),

$$B_j(L,D) = 2\frac{v_F m_e^*}{e}\frac{D}{\left(\frac{L}{2j}\right)^2 + D^2}\left[\cos(\phi) - \frac{L}{2jD}\sin(\phi)\right] \qquad \text{Eq. (1).}$$

In previous works[9,11] given the symmetry of the adopted cavities, it was assumed that electrons enter the cavity aligned with its symmetry axis ($\phi=0$) and the dependence $B_j(\phi)$ was neglected. In our case, given the shape of the cavities, we had to consider the $\phi$ dependence because no preferential value for the entrance angle $\phi$ is present. As can be se in Eq. (1), the $\phi$ dependence appears as a factor $f_j(\phi,\alpha) = [\cos(\phi) - (\gamma/2j)\sin(\phi)]$ that depends on the aspect ratio of the cavity $\gamma = L/D$ and can influence substantially the value of $B_j$. In comparing the model with the experimental data we used $\phi$ as a free parameter in order to best fit the experimental values $B_j$ of the magneto-conductance maxima with the theoretical ones calculated according to Eq. (1). As an example, in Fig. 5 we report a magneto-conductance curve taken at gate voltage Vg=0 V where magneto-conductance oscillations are measured in an extended range and we show with arrows the first three magneto-conductance maxima values $B_j$ calculated according to Eq. (1). The reported $B_j$ ($j = 1, 2, 3$) were calculated using the measured values of the geometrical parameters of the cavity (L, D) and the physical parameters ($v_F$, $m_e^*$) of the 2DEG and using the entrance angle $\phi$ as a best-fit parameter. The best agreement between the experimental and the theoretical $B_j$ was obtained for $\phi=11.3°$. We report the experimental and theoretical values of $B_j$ as an inset in Fig. 5. Although in a more accurate analysis the experimental magneto-conductance should be compared with a semiclassical simulated curve based on billiard-ball models,[32] the good agreement between experimental values and the results of the simple geometrical model illustrated above lend support to the interpretation in terms of magnetic focusing.

### C. Weak localization

We will now discuss the zero-field peak in the magnetoresistance curves and the features close to it at low magnetic fields (B<0.1 T). We have seen in Fig. 3 that two kinds of peaks are present at B=0. In the $R_{1,2}$ curve the peak is sharp and narrow (height in the order of 1 kΩ and full width at half-maximum ~ 8 mT), while in the $R_{2,3}$ curve is weaker and broad. The structure in the $R_{1,4}$ curve is clearly the superposition of the two peaks. Remembering that in the $R_{1,2}$ curve the current flow outside the cavity along a path whose width is always larger than $L_\Phi$ and that only the current flowing inside the cavity contribute to $R_{2,3}$, we conclude that the narrow peak is due to disorder-induced weak localization in 2D diffusive regime, whereas the broad peak is interpreted in term of "ballistic weak localization". Clearly, both effects contribute to the two-terminal magnetoresistance $R_{1,4}(B)$.

From the width of the narrow peak as measured at 50 mK we have determined $B_C=3.7$ mT, i.e. a value practically coincident with that in the Hall bar reported above.

The four-terminal $R_{23}$ data were investigated at T=50 mK as a function of the gate voltage in the open transport regime. In Fig. 6(a) we report a series of 11 magnetoresistance curves measured in sequence, progressively decreasing the gate voltage from +1 V to -1 V in steps of 0.2 V. At first glance the curves seem to be quite different each other for what concerns the peak lineshape. As an example, the curve measured at Vg=+0.6 V seems to have a linear shape, rather different from the more rounded Lorentzian-like peaks found at Vg= -1 V, -0.2 V or +0.4 V. Noteworthy is the curve at Vg=+0.8 V in which the zero-field peak has almost disappeared in a flat plateau. Noticeable features are also the shoulders that develop around ±50mT (marked with dashed lines as guide for the eyes) and increase or decrease non monotonously as the gate voltage is changed. We shall return to this point below.

In order to perform a preliminary lineshape analysis, all the curves of Fig. 6(a) have been fitted in the (-100, +100) mT range by the Lorentzian lineshape predicted for ballistic WL in a chaotic cavity

$$R(B) = R_0 + \frac{\Delta}{1 + (2B/\alpha\Phi_0)^2} \qquad \text{Eq. (2)}$$

where $\Phi_0 = (h/e)$ is the flux quantum, $\Delta = (R(B=0) - R_0)$ is the zero-field WL correction and $\alpha$ the inverse of the typical area enclosed by the classical trajectory of the ballistic electron in the cavity. In panel (b), (c), and (d) of Fig. 6 we report some representatives curves, selected from those of panel (a), together with the Lorentzian fit performed according to Eq. (2) and using $\alpha$ and $R_0$ as fitting parameters. Looking at the curve in panel (b) (Vg=-1 V), it is evident that the Lorentzian fit agrees quite well with the experimental data in the central region, but deviates at higher magnetic fields because of the shoulders, marked with arrows in figure, that rise at B = ±74mT. Panel (c) highlights the situation when a Lorentzian curve fits well the experimental data in an extended range of magnetic field (up to B=0.1 T). In panel (d) we report the case of the triangular-shaped peak found at Vg=+0.6 V. We see that, due to the presence of a somewhat rounded top and barely noticeable shoulders at +/- 50 mT (marked with arrows), it can also be satisfactorily fitted with a Lorentzian lineshape in the central region. In Fig. 6(e) we summarize the results of the Lorentzian fit of all the magnetoresistance curves, reporting the quantity $\alpha\Phi_0$ as a function of the gate voltage. The extracted $\alpha\Phi_0$ do not vary significantly with the gate voltage and this can be interpreted as an indication that changes in the gate voltage result in a variation of the carrier density in the cavity without altering significantly its shape. This is somehow expected in our devices since they are fabricated with etching rather than split-gate techniques.

Averaging the $\alpha\Phi_0$ obtained from the Lorentzian fits we obtain $\alpha\Phi_0^*=(51.5\pm1.8)$ mT, from which we can estimate the typical area enclosed by the classical electron trajectories $\alpha^{-1}=0.08\mu m^2$ in close agreement with the value $A=0.11\mu m^2$ of the area of the cavity estimated from the lithographic dimensions. This agreement was also found in previous studies of WL in ballistic cavities realized by split gates on GaAs 2DEGs.[16,17]

As a conclusion of the data inspection and the above preliminary analysis, we point out the close similarity between our results and previous findings on a variety of cavities realized from GaAs-based 2DEGs.[16,17] The central portion of the spectra could be fitted, but important discrepancies were found on the side of the B=0 peak. The fitting parameter gives a reasonable estimate of the cavity area. The analysis was based on the theory,[12,13] which attributes the peaks to weak localization arising from interference between backscattered trajectories. Furthermore, this theory predicts a Lorentzian lineshape for chaotic trajectories and a linear dependence for regular ones. Given the overall discrepancies between data and theory and the puzzling observation of linear and Lorentzian lineshape in the same cavities, it was argued, from one side, that a large ensemble average is necessary in order to compare correctly the experimental results,[12,13] on the other side, it was suggested that zero-field peak should be attributed to a phenomenon other than weak localization.[18,19]

In order to check these possibilities, we have refined our analysis in the WL framework by taking into account the fact that the theoretical lineshapes are calculated by averaging over all wave vector $k$ and thus the correct quantities to compare with the theory are energy-averaged quantities. In our case, assuming that the gate voltage varies the carrier density only and so the wave vector $k$, we have performed an energy-averaged analysis similar to that reported in Ref. 33. The WL energy-averaged conductance correction $<\Delta G(B)>_k$ was calculated according to

$$<\Delta G(B)>_k = \frac{1}{M}\sum_{i=1}^{M} G(B,V_G(i)) - G(B=0,V_G(i)) \qquad \text{Eq. (3)},$$

where the magneto-conductance curves $G(B,V_G)$ were derived from the magnetoresistance measurements reported in Fig. 6(a) and the summation index $i$ refers to data extracted from the M=11 curves measured at different gate voltage values. The experimental points were then fitted with the inverted Lorentzian-like form predicted for ballistic WL in chaotic cavities:[12,13,34]

$$<\Delta G(B)>_k = \Sigma\left[1 - \frac{1}{1+(2B/\alpha\Phi_0)^2}\right] \qquad \text{Eq. (4)}$$

where the amplitude $\Sigma$ related to the number of modes in the leads and the number of effective phase breaking channels, at zero temperature, is predicted to assume the value $\Sigma = 0.5 e^2/h$.

In Fig. 7 we report as solid squares the experimental energy averaged WL effect $<\Delta G(B)>_k$ as a function of the magnetic field expressed in unity of $\alpha\Phi_0$, assuming for this the value $\alpha\Phi_0 = \alpha\Phi_0^* = 51.5$ mT estimated previously. For clarity the corresponding magnetic field B is also reported. For each experimental point, the error bars extracted from the averaging procedure are also reported. In the magnetic field range where the $<\Delta G(B)>_k$ develops the typical WL dip, the error bars are noticeably small. At higher magnetic fields, as remarked before, the magnetoresistance curves are quite different from each other and this, together with the fact that a limited number of curves was used for averaging, leads to progressively increasing error bars.

The solid line is the fit of Eq. (4) to the experimental points, using the amplitude $\Sigma$ as the fitting parameter and fixing, once again, $\alpha\Phi_0 = 51.5$ mT. The value $\Sigma = 0.16 e^2/h$ gives an excellent agreement of the fitting curve with the experimental data in the (-40, +40) mT range. Amplitude values $\Sigma$ lower than the theoretically predicted ones are reported in literature for cavities of different shape and related to the existence of short paths for electrons in the cavity.[33] Referring to Fig. 7, clear deviations of the fitting curve from the experimental data are observed at higher magnetic fields, especially in the range where the $<\Delta G(B)>_k$ curve develops smooth kinks (highlighted with arrows) that are the reflection of the side-peak shoulders in the magnetoresistance curves and that remain after the curve averaging. Therefore, the present analysis confirms the previous conclusions: the central portion of the spectra could be fitted, but important discrepancies remain on the side of the B=0 peak that the ballistic weak localization theory cannot account for.

As mentioned previously, a similar situation found in GaAs-based cavities was the starting point of the theoretical approach developed by Akis and co-workers,[18,19] claiming that the conductance displays features reflecting the energy spectrum of a ballistic cavity. Unfortunately, in this case no simple theoretical equations are available for data analysis. However, we emphasize the close similarity between our cavities and the geometry used in the simulations of Ref. 18, 19 and the striking resemblance between our results and their findings (in particular, compare our Fig. 7 with Fig. 7 of Ref. 19). We conclude, therefore, that our results are in line with the analysis of Akis et al. in which the zero-field peak and the related side features in the magnetoresistance are viewed as a reflection of the intrinsic energy spectrum of the ballistic quantum dot.

## IV. CONCLUSIONS

In conclusion, we have investigated Si/SiGe-based ballistic cavities for the first time. The adopted geometry was such that the carrier density and the "isolation" of the central dot could be varied by the gate bias. The magnetoresistance curves exhibit intensity modulations, that we have explained in terms of electron focusing effects. A zero-magnetic-field peak was also present in the spectra. A careful inspection of the peak lineshape revealed the presence of side-peak shoulders in the magnetoresistance curve that have no explanation in the framework of the ballistic weak localization theory. The side-peak features survive to energy averaging and can be viewed as a reflection of the intrinsic spectrum of the ballistic quantum dot, as suggested in previous studies on similar GaAs-based cavities.

## REFERENCES

[1]H. van Houten, C. W. J. Beenakker, J. G. Williamson, M. E. I. Broekaart, P. H. M. van Loosdrecht, B. J. van Wees, J. E. Mooij, C. T. Foxon, and J. J. Harris, Phys. Rev. B **39**, 8556 (1989).

[2]B. J. van Wees, H. van Houten, C. W. J. Beenakker, J. G. Williamson, L. P. Kouwenhoven, D. van der Marel, and C. T. Foxon, Phys. Rev. Lett. **60**, 848 (1988).

[3]V. J. Goldman, B. Su, and J. K. Jain, Phys. Rev. Lett. **72**, 2065 (1994).

[4]J. H. Smet, D. Weiss, R. H. Blick, G. Lutjering, K. von Klitzing, R. Fleischmann, R. Ketzmerick, T. Geisel, and G. Weimann, Phys. Rev. Lett. **77**, 2272 (1996).

[5]R. M. Potok, J.A. Folk, C. M. Marcus, and V. Umansky, Phys. Rev. Lett. **89**, 266602 (2002).

[6]L. P. Rokhinson, V. Larkina, Y. B. Lyanda-Geller, L. N. Pfeiffer and K. W. West, Phys. Rev. Lett. **93**, 146601 (2004).

[7]R. P. Taylor, A. S. Sachrajda, J. A. Adams, P. T. Coleridge and P. Zawadzki, Phys. Rev. B **47**, 4458 (1993).

[8]T. M. Eiles, J. A. Simmons, M. E. Sherwin and J. F. Klem, Phys. Rev. B **52**, 10756 (1995).

[9]Y. Ochiai, A. W. Widjaja, N. Sasaki, K. Yamamoto, R. Akis, D. K. Ferry, J. P. Bird, K. Ishibashi, Y. Aoyagi and T. Sugano, Phys. Rev. B **56**, 1073 (1997).

[10]P. Boggild, A. Kristensen and P. E. Lindelof, Phys. Rev. B **59**, 13067 (1999).

[11]P. D. Ye and S. Tarucha, Phys. Rev. B **59**, 9794 (1999).

[12]H. U. Baranger, R. A. Jalabert and A. D. Stone, Phys. Rev. Lett. **70**, 3876 (1993).

[13]H. U. Baranger, R. A. Jalabert and A. D. Stone, Chaos **3**, 665 (1993).

[14]A. M. Chang, H. U. Baranger, L. N. Pfeiffer and K. W. West, Phys. Rev. Lett. **73**, 2111 (1994).

[15]M. J. Berry, J. A. Katine, R. M. Westervelt and A. C. Gossard, Phys. Rev. B **50**, R17721 (1994).

[16]Y. Lee, G. Faini and D. Mailly, Phys. Rev. B **56**, 9805 (1997).

[17]J. P. Bird, D. M. Olatona, R. Newbury, R. P. Taylor, K. Ishibashi, M. Stopa, Y. Aoyagi, T. Sugano and Y. Ochiai, Phys. Rev. B **52**, R14336 (1995).

[18]R. Akis, D. Vasileska, D. K. Ferry and J. P. Bird, J. Phys. Condens. Matter **11**, 4657 (1999).

[19]R. Akis, D. K. Ferry, J. P. Bird and D. Vasileska, Phys. Rev. B **60**, 2680 (1999).

[20]J. P. Bird, R. Akis and D. K. Ferry, Phys. Rev. B **60**, 13676 (1999).

[21]J. P. Bird, R. Akis, D. K. Ferry, A. P. S. de Moura, Y-C. Lai, K. M. Indlekofer, Rep. Prog. Phys. **66**, 583 (2003).

[22]A. Notargiacomo, L. Di Gaspare, G. Scappucci, G. Mariottini, F. Evangelisti, E. Giovine and R. Leoni, Appl. Phys. Lett. **83**, 302 (2003).

[23]L. Di Gaspare, K. Alfaramawi, F. Evangelisti, E. Palange, G. Barucca and G. Majni, Appl. Phys. Lett. **79**, 2031 (2001).

[24]L. Di Gaspare, G. Scappucci, E. Palange, K. Alfaramawi, F. Evangelisti, G. Barucca and G. Majni, Mater. Sci. Eng. B **89**, 346 (2002).

[25]P. T. Coleridge, R. Stoner and R. Fletcher, Phys. Rev. B **39**, 1120 (1989).

[26]D. Tobben, F. Schaffler, A. Zrenner, and G. Abstreiter, Phys. Rev. B **46**, R4344 (1992).

[27]P. T. Coleridge, Phys. Rev. B **44**, 3793 (1991).

[28]R. S. Prasad, T. J. Thornton, A. Matsumura, J. M. Fernandez and D. Williams, Semicond. Sci. Technol. **10**, 1084 (1995).

[29]G. Stoger, G. Brunthaler, G. Bauer, K. Ismail, B. S. Meyerson, J. Lutz and F. Kuchar Semicond. Sci. Technol. **9**, 765 (1994).

[30]L.-H. Lina, N. Aoki, K. Nakao, K. Ishibashi, Y. Aoyagi, T. Sugano, N. Holmberg, D. Vasileska, R. Akis, J. P. Bird, D. K. Ferry and Y. Ochiai, Physica E **7**, 750 (2000).

[31]R. Brunner, R. Meisels, F. Kuchar, M. El Hassan, J. Bird and K. Ishibashi, Physica E **21**, 491 (2004).

[32]H. Linke, L. Christensson, P. Omling and P. E. Lindelof, Phys. Rev. B **56**, 1440 (1997).

[33]M. W. Keller, A. Mittal, J. W. Sleight, R. G. Wheeler, D. E. Prober, R. N. Sacks and H. Shtrikmann, Phys. Rev. B **53**, R1693 (1996).

[34]R. A. Jalabert, H. U. Baranger and A. D. Stone, Phys. Rev. Lett. **65**, 2442 (1990).

**Table I**

| | Si/SiGe 2DEG parameters |
|---|---|
| $n_{2D}$ | $7.62 \times 10^{11} \text{cm}^{-2}$ |
| $\tau_E$ | $7.3 \times 10^{-12}$ s |
| $\mu_E$ | $6.7 \times 10^4$ cm$^2$/Vs |
| $\lambda_E$ | 0.7 µm |
| $D$ | 320 cm$^2$/Vs |
| $v_F$ | $9.4 \times 10^4$ m/s |
| $\tau_q$ | $2.3 \times 10^{-12}$ s |
| $\mu_q$ | $2.1 \times 10^4$ cm$^2$/Vs |
| $\tau_E/\tau_q$ | 3.2 |
| $\tau_\Phi$ | $\sim 7 \times 10^{-12}$ s |
| $L_\Phi$ | $\sim 0.5$ µm |

Table I. Summary of the low temperature (T=250 mK) transport parameters of the Si/SiGe 2DEG determined experimentally from the low-field magnetoresistance of mesa-etched Hall bars.

**Figure 1**

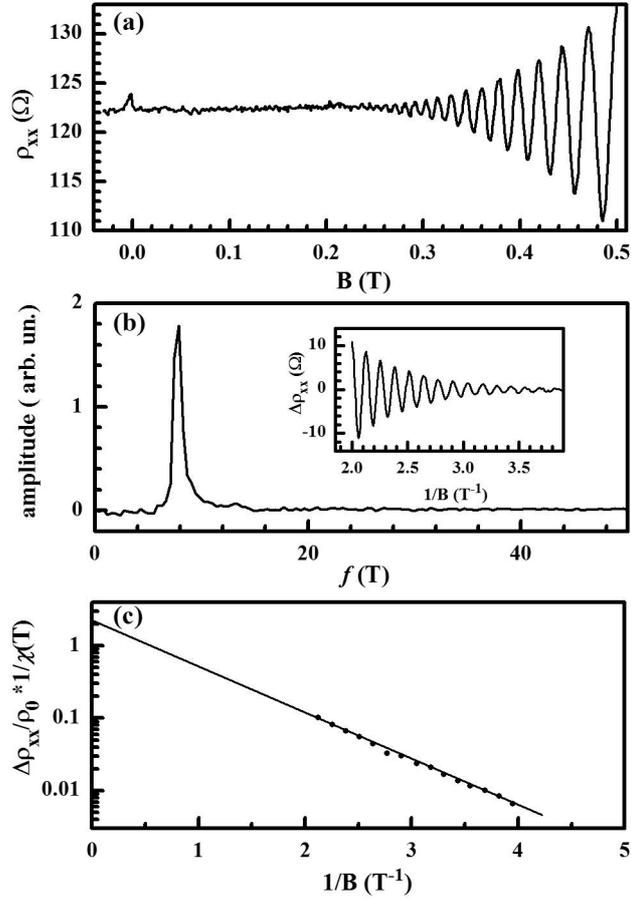

FIG. 1. (a) Low-field magnetoresistivity $\rho_{xx}(B)$ of mesa-etched Hall bars (T=250 mK) displaying a zero-field diffusive weak localization peak and Shubnikov-de Haas oscillations. (b) Fourier spectrum of the oscillatory part $\Delta\rho_{xx}(B)$, shown in the inset vs. the inverse magnetic field. (c) Reduced resistivity $\Delta\rho_{xx}/\rho_0$, corrected by the temperature damping factor $\chi(T,\omega_C)$ as defined in the text, vs. the inverse magnetic field. The straight line is the one parameter linear fit used to estimate the single particle lifetime.

**Figure 2**

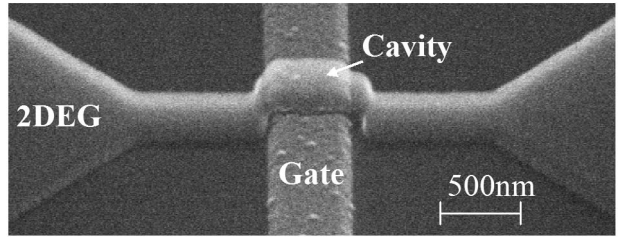

FIG. 2. Scanning electron micrograph of a cavity fabricated by reactive ion etching of a Si/SiGe modulation doped heterostructure.

**Figure 3**

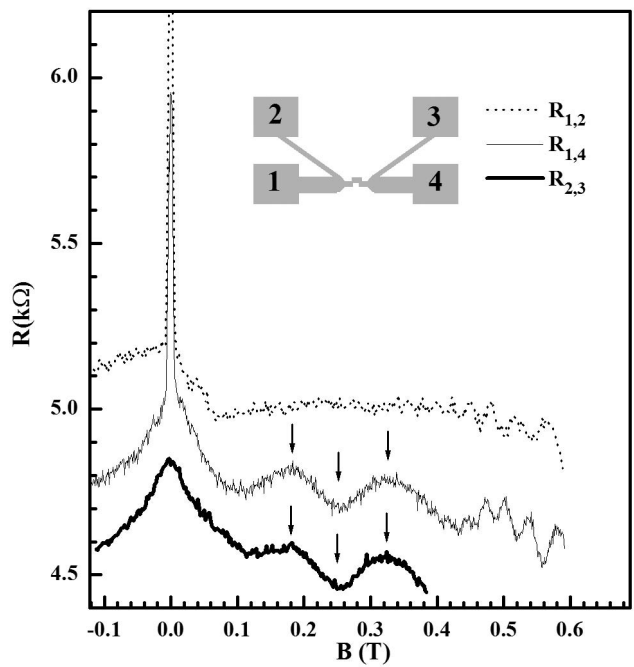

FIG. 3. Low field magnetoresistance at T=50 mK measured through the cavity in the four-terminal configuration ($R_{2,3}$, thicker line) and in the two-terminal configuration ($R_{1,4}$, thinner line) as well as a two-terminal "background" resistance ($R_{1,2}$, dotted line). Electrical contacts schematics is reported in the inset.

**Figure 4**

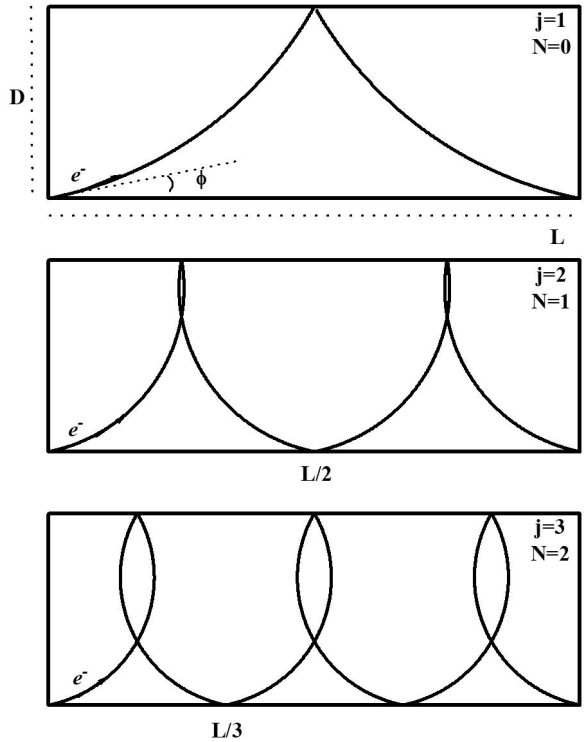

FIG. 4. Classical electron trajectories in a rectangular cavity that lead to maxima in the magneto-conductance. The commensurability condition are met when the electron, entering the cavity with an angle $\phi$ (sizes L x D) on the lower left corner, first collide with the lower wall at a fraction $1/j$ of its length and leaves the cavity after $N = j - 1$ collisions with the lower wall.

**Figure 5**

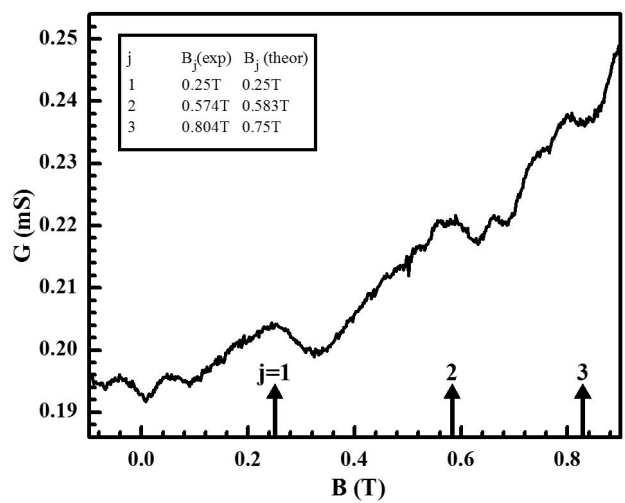

FIG. 5. Magneto-conductance oscillations measured in an extended range at T=50mK. The arrows correspond to the first three magneto-conductance maxima positions $B_j$ calculated using a semiclassical billiard model with the entrance angle $\phi$ used as a fit parameter. In the inset a comparison between experimental peak positions and calculated ones is reported.

**Figure 6**

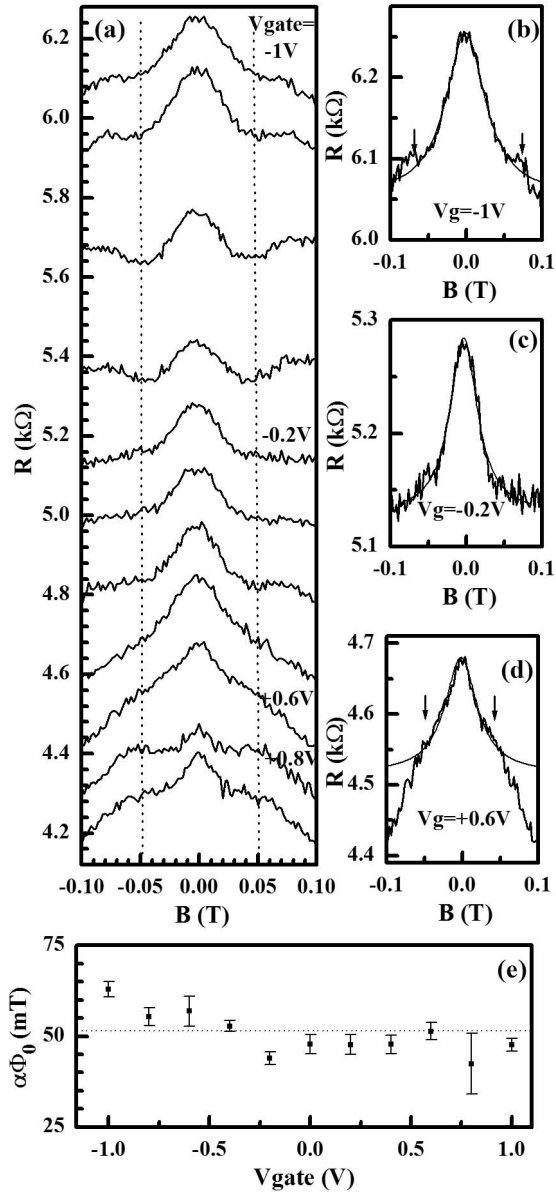

FIG. 6. (a) Four-terminal magnetoresistance investigated as a function the gate voltage, decreased progressively from +1V to -1V in steps of 0.2V. Some representative curves are reported in panel (b), (c) and (d) together with a lorenztian fit according to Eq. (2) using $\alpha$ and $R_0$ as fitting parameters. A good agreement is obtained only in the central region of the peaks. The obtained $\alpha\Phi_0$ for all curves shown in panel (a) are reported vs. the gate voltage in panel (e).

**Figure 7**

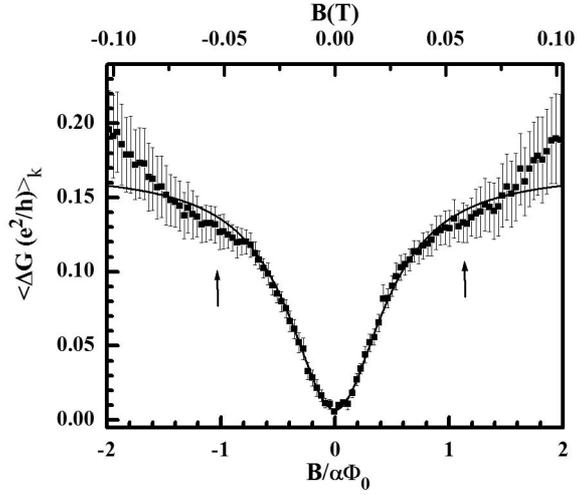

FIG. 7. Experimental energy averaged weak localization effect as a function of magnetic field expressed in unity of $\alpha\Phi_0$ with $\alpha\Phi_0=51.5$mT. The solid line is the theoretical fit to the experimental data according to Eq (4) using the amplitude $\Sigma$ as fitting parameter. Clear discrepancies between experimental points and theoretical fit are marked with arrows.